\documentclass[preprint,prb,amsmath,groupaddress,aps,subeqn,showpacs]{revtex4}
\usepackage{graphicx}

\begin{document}

\title{Strong Substrate Dependence of Joule Heating in Graphene}

\author{X. Li}
\affiliation{Department of Electrical and Computer Engineering, North Carolina
State University, Raleigh, NC 27695-7911}

\author{B. D. Kong}
\affiliation{Department of Electrical and Computer Engineering, North
Carolina State University, Raleigh, NC 27695-7911}

\author{J. M. Zavada}
\affiliation{Department of Electrical and Computer Engineering, North
Carolina State University, Raleigh, NC 27695-7911}

\author{K. W. Kim}
\affiliation{Department of Electrical and Computer Engineering, North Carolina
State University, Raleigh, NC 27695-7911}

\begin{abstract}

The Joule heating effect on graphene electronic properties is
investigated by using full-band Monte Carlo electron dynamics and
three-dimensional heat transfer simulations self-consistently.  A
number of technologically important substrate materials are
examined: SiO$_2$, SiC, hexagonal BN, and diamond.  The results
illustrate that the choice of substrate has a major impact via the
heat conduction and surface polar phonon scattering.  Particularly,
it is found that the poor thermal conductivity of SiO$_2$ leads to
significant Joule heating and saturation velocity degradation in
graphene (characterized by the so-called 1/$\sqrt{n}$ decay).
Considering the overall characteristics, BN appears to compare
favorably against other substrate choices for graphene in electronic
applications.

\end{abstract}


\maketitle

Superior electronic properties are one of the main attractions of graphene in device applications.  The massless Dirac Fermions originating from a linear dispersion relation imply  high electron mobilities and drift velocities $-$ an ideal trait for devices in high-frequency/high-speed operation.~\cite{Avouris2010}  Nevertheless, carrier motions are subject to scattering by perturbations such as lattice vibrations or impurities in most (if not, all) realistic conditions. Particularly, the electron interaction with the lattice is essentially a thermalization process from a system point of view.  As electrons gain energy from an external source (such as an electrical bias), a part of the excess energy is transferred to the lattice via phonon emission.  Subsequent increase in the lattice temperature (i.e., the Joule heating) acts as a counter weight to limit further energy gain from the source by causing degradation in the electronic transport. Eventually, a balance is reached and the system approaches the steady state.  Thus, the details of heat dissipation including the properties of its primary path (i.e., the substrate)  could have a major influence.  This is even more so in graphene based structures,~\cite{Liao2011} where the two-dimensional (2D) nature dictates a large interface with the substrate compared to the volume.

In this Letter, we theoretically investigate the effect of Joule heating in graphene.  Specifically, the impact of different substrates on graphene electron transport properties are examined with four technologically important substrate materials: SiO$_2$, SiC, hexagonal BN (\textit{h}-BN), and diamond.  While SiO$_2$ is the most commonly used substrate due to its compatibility with conventional technology,~\cite{Meric2008,Dorgan2010} SiC has seen its use in producing large-area graphene by solid state graphitization.~\cite{Hass2006}  Recently, \textit{h}-BN has drawn attention for its structural similarity to graphene $-$ a much desired condition for high quality samples.~\cite{Xue2011} On the other hand, diamond can also be a candidate.  Beside the anticipated affinity with graphene, it is one of the most thermally conductive materials. In the analysis, we consider a model problem, where a small graphene sample is placed on a 2D plane of relatively thick dielectric or substrate (300 nm), which is in turn on top of a bulk Si layer.  The graphene film is subject to a uniform electric field, generating excess heat that must be dispersed through the layers underneath.  To obtain the self-consistent solution across the structure, we solve simultaneously the 3D heat transfer equation (including the estimated interfacial thermal resistance) and the Monte Carlo electron dynamics in graphene.

The Monte Carlo simulation developed in this study takes into
account the complete electron and phonon spectra in the first
Brillouin zone.  Specifically, both the graphene phonon dispersion
and its interaction with electrons are obtained from the
\textit{density functional theory}
calculations,~\cite{Borysenko2010} whereas a tight binding model is
used for the electronic energy bands.~\cite{Neto2009}  In addition,
graphene electron interactions with charged impurities
($5\times10^{11}$~cm$^{-2}$) on the substrate
surface~\cite{Adam2009} as well as the surface polar phonons (SPPs)
are included.~\cite{Perebeinos2010} For simplicity, the phonon
system is assumed to reach thermal equilibrium fast enough so that
the electron-phonon scattering rates, for both intrinsic graphene
phonons and SPPs, have the temperature dependence based on the
Bose-Einstein distribution: i.e., $N_q=1/[exp(\frac{\hbar
\omega_{ph}} {k_BT})-1]$, where $\omega_{ph}$ is the phonon
frequency, $k_B$ the Boltzmann constant, and $T$ the temperature.

The electron energy transferred to the lattice vibrational modes increases the lattice temperatures of graphene and the substrate.  Specifically, the interaction with graphene phonons (e.g., emission) leads to the elevation of graphene lattice temperature ($T_g$), whereas the substrate has contributions from both the direct excitation of SPPs (i.e., electron-SPP scattering) and the heat conduction from the graphene lattice.  As summarized above, the thermal part of the self-consistent model utilizes a 3D heat transfer equation in the substrate; $ \nabla \cdot [\kappa(x,y,z) \nabla T(x,y,z)] =0 $, where $\kappa$ is the thermal conductivity of the material.  The values used in the calculations are $\kappa_{\rm SiO_2} =1.4$~Wm$^{-1}$K$^{-1}$, $\kappa_{\rm SiC} =370 $~Wm$^{-1}$K$^{-1}$ and $\kappa_{\rm dia} =1800$~Wm$^{-1}$K$^{-1}$ for SiO$_2$, SiC, and diamond, respectively.~\cite{handbook,Worner1996}  Unlike the first three, the thermal conductivity of \textit{h}-BN is anisotropic with a large difference in the in-plane and the out-of-plane direction due to the layered nature; $\kappa_{h\textrm{-BN}}(x,y)\approx300$~Wm$^{-1}$K$^{-1}$ and $\kappa_{h\textrm{-BN}}(z)\approx 2 $~Wm$^{-1}$K$^{-1}$.~\cite{Simpson1971}  The corresponding details on Si can be found in Ref.~\onlinecite{Glassbrenner1964}.

In a heterogeneous system, there is an extra thermal resistance $r_{gs}$ at
the interface between graphene and the substrate (i.e., the so-called Kapitza resistance).  An experimental measurement  reported $r_{gs}$ ranging from  $5.6\times10^{-9}$ to $1.2 \times 10^{-8}$~Km$^2$W$^{-1}$ in the graphene/SiO$_2$ structure.~\cite{Chen2009}  Interestingly, a first principles calculation conducted very recently also suggests similar numbers for the graphene interface with \textit{h}-BN and SiC.~\cite{Mao2011}  As such, a typical value of $8.8\times10^{-9}$~Km$^2$W$^{-1}$ (i.e., the median of the range observed for SiO$_2$) is adopted in this study for all four substrate materials.  Concerning SiC, however, the situation is more complex.  When it is exposed to the air, hydrogen tends to be adsorbed and terminate the dangling bonds of Si lonely atoms in order to form a stable surface.~\cite{Jayasekera2010}  The surface is also passivated intentionally to reduce the interface states.  As there is an indication that this could cause a drastic increase in $r_{gs}$,~\cite{Mao2011} an additional case of fully hydrogen terminated SiC (SiC-H) is considered with $r_{gs} = 7.9 \times 10^{-8}$~Km$^2$W$^{-1}$ to gauge the impact. In the real situation, the surface is more likely to show partial termination (i.e., somewhere between the cases of SiC and SiC-H).

Then, the total power dissipation per unit area across the interface between graphene and the substrate can be expressed as $P_{t}=(T_g-T_s)/r_{gs}+P_{spp}$, where $T_s$ is the lattice temperature of the substrate at the interface and $P_{spp}$ is the power transferred via the direct SPP scattering.  As $P_t$ should be equal to the net power loss by graphene electrons in a steady state, both this quantity and $P_{spp}$ can be estimated from the Monte Carlo simulation, providing the necessary boundary condition for the heat transfer equation.  Finally, a self-consistent temperature profile (including $T_g$ and $T_s$) is obtained by an iterative process.

Figure~1 illustrates the potential impact of Joule heating in graphene on   different substrate materials.  Two sets of data are provided with the electron density of $1 \times10^{12}$~cm$^{-2}$ and the graphene sample dimension of 1~$\mu$m $ \times$ 0.5~$\mu$m.  One set of results, shown in lines, represents the drift velocity versus electric field when Joule heating is ignored by fixing $T_g$=$T_s$=300~K.  The other set, in data points, examines the same curves with the Joule heating effect taken into account.  Of the cases under consideration, the most drastic changes appear in the graphene/SiO$_2$ structure.   While the deviation is minor in the low-field region, the saturation velocity $v_{sat}$ degrades substantially (e.g., by about 20\% to $3.9\times10^7$~cm/s at 30 kV/cm).  In comparison, the velocity-field curves show little impact of Joule heating for all other substrates.  Only SiC-H shows a minor influence; the rest including SiC (unpassivated; not shown in Fig.~1) appear virtually unaffected.  Here, it is also interesting to note that the graphene-on-diamond structure actually has the lowest $v_{sat}$ independent of Joule heating.  The origin of this departure is the absence of SPPs in diamond.  Without the SPP scattering, the electrons in graphene lose a major energy relaxation mechanism and stay hotter than otherwise. Consequently, it is not unreasonable to expect a smaller drift velocity on a non-polar substrate (such as diamond) than on a polar counterpart.~\cite{Li2010}  Indeed, the results show that the average graphene electron energy on the diamond substrate is $0.44$~eV at $30$~kV/cm, while it is only $0.25$~eV on SiO$_2$.

To better examine the observed impact on electron drift velocities, the temperatures in the graphene film ($T_g$) and at the interface directly below ($T_s$) are provided in Fig.~2 as a function of applied bias. Both $T_g$ and $T_s$ increase with the field but their magnitudes vary widely depending on the substrate.  Clearly, SiO$_2$ shows the extreme case of Joule heating with very high $T_g$ and $T_s$ that is consistent with the results of Fig.~1.  Due to the poor thermal conductivity $\kappa_{\rm SiO_2}$, the excess heat emitted by graphene electrons cannot be efficiently channeled through the substrate.  The resulting increase in temperature induces stronger electron-phonon scattering and subsequently degrades the drift velocity.  As for \textit{h}-BN, the rise in $T_g$ and $T_s$ is far more modest despite the very low out-of-plane thermal conductivity $\kappa_{h\textrm{-BN}}(z)$, which is in fact about the same order of magnitude as $\kappa_{\rm SiO_2}$.  The discrepancy comes from  the in-plane thermal conductivity $\kappa_{h\textrm{-BN}}(x,y)$ that is about two orders of magnitude larger.  Consequently, the transferred heat in \textit{h}-BN can easily spread in-plane unlike in SiO$_2$, utilizing a much wider thermal channel. Indeed, the 3D iso-temperature profile illustrates a laterally extended distribution near the interface $-$ a sign of efficient heat removal.  Two substrates with high thermal conductivities, diamond and SiC (unpassivated; not shown), show even smaller deviations from room temperature as expected.

In the case of SiC-H, the moderate increase in $T_g$ (and the subsequent velocity decay) has a different origin.  While thermal transport in the substrate is excellent (owing to a superior $\kappa_{\rm SiC}$), the heat conduction across the interface with a relatively large $r_{gs}$ provides the bottleneck.  This point is clearly illustrated in Fig.~2(b) by the largest $T_g-T_s$ among those plotted; the substrate surface temperature $T_s$ stays near 300~K in the graphene/SiC-H structure.  Another interesting observation in Fig.~2(b) is that the absence of SPP interaction is visible from the result of diamond.  The comparatively large $T_g-T_s$ (over those of \textit{h}-BN and SiO$_2$) indicates a greater disconnect between the graphene film and the substrate in terms of heat transfer.  Since an identical $r_{gs}$ is assumed (except SiC-H), the absence of additional heat path via SPP emission appears to be the main reason for the increased $T_g-T_s$.

Finally, an attempt is made to compare the simulation results with the experimental studies available in the literature. Since the detailed heat dissipation (thus, the extent of Joule heating) depends on such factors as the exact geometry, etc., that not only vary from sample to sample but also are not fully characterized, it is rather difficult to have a meaningful one-to-one comparison at the quantitative level.  For instance, when the lateral dimension of the graphene film is much larger than the thickness of substrate dielectric,~\cite{Liao2011} heat transfer through the structure could simply be projected to a 1D problem with the outcome potentially much different from that modeled in the current calculation.  The 3D effect such as the lateral heat spread would be absent and \textit{h}-BN would behave more like SiO$_2$ with pronounced Joule heating.  At the same time, the additional structural details including the location and the dimension of metal contacts could alter the heat dissipation pattern that are not included in the current model.  Consequently, we treat the graphene sample size as an effective parameter that is adjusted to provide a good fit with the experimental data (specifically, those on SiO$_2$).~\cite{Meric2008,Dorgan2010}

Figure~3 shows the comparison of $v_{sat}$ as a function of electron density $n$.  The results clearly indicate the increased prominence of Joule heating at large values of $n$.  This is obvious as more electrons mean lager heat generation per unit area, which leads to elevated lattice temperature and reduced $v_{sat}$.   Consistent with the results discussed above, SiO$_2$ suffers the biggest impact and then SiC-H is the next while the rest (BN, SiC, and diamond) remain largely unaffected in the considered range ($n \lesssim 4 \times 10^{12}$~cm$^{-2}$).  One particularly interesting point to note is the slope of decay.  The slope deduced from the simulation appears to become steeper in a close correlation with the rise in $T_g$ and reaches the $1/\sqrt{n}$ dependence for SiO$_2$ that matches well with the experimental data from Refs.~\onlinecite{Meric2008} and \onlinecite{Dorgan2010}. This result, however, cannot be explained when the Joule heating is excluded as evident from the figure (see the solid line).  Consequently, it strongly indicates that the so-called $1/\sqrt{n}$ decay does not come from the SPP energy of the SiO$_2$ substrate as originally suggested.~\cite{Meric2008}  Rather, it is a manifestation of Joule heating in its entirety including the influence of SPP scattering characteristics $-$ a case-specific outcome and not a general rule.

Considering the results thus far, SiO$_2$ may not be a desirable choice as a substrate material.  As for diamond, the lack of SPP scattering appears to result in the drift velocities substantially smaller than other candidates (such as BN and SiC).  However, it is also the most immune from the Joule heating degradation for its high thermal conductivity and may have an advantage at very high carrier densities ($\gtrsim 10^{13}$~cm$^{-2}$).  While the performance of BN and SiC are generally comparable, SiC-H shows the sign of elevated temperatures in graphene.  This could cause a significant concern in device breakdown characteristics in addition to the channel velocity reduction. Since the dangling bonds at the surface tend to be terminated in one form or another (by hydrogen or other specifies), the realistic structure involving the SiC substrate may be more like SiC-H than the ideal case without termination.
Thus, it appears that \textit{h}-BN provides the best characteristics among the studied to interface with graphene in electronic applications.

This work was supported in part by the US ARO, DARPA/HRL CERA, and NERC/NIST SWAN-NRI programs. JMZ acknowledges support from NSF under the IR/D program.

\clearpage
\noindent Figure Captions

\vspace{0.5cm} \noindent Figure~1. (Color online) Drift velocities
vs.\ electric field for graphene on different substrates.  The
graphene sample is assumed to be 1~$\mu$m $ \times$ 0.5~$\mu$m with a carrier density of $1\times 10^{12}$~cm$^{-2}$. The impurity density is $5 \times
10^{11}$~cm$^{-2}$.

\vspace{0.5cm} \noindent Figure~2. (Color online) (a) Graphene
lattice temperature $T_g$ for different substrates, and (b)
temperature difference $T_g-T_s$ between the graphene lattice and the top
surface of the substrate as a function of driving electric field.  The conditions are the same as in Fig.~1.

\vspace{0.5cm} \noindent Figure~3. (Color online) Saturation velocity $v_{sat}$ vs.\ electron density in graphene on different substrates.  In the calculations, it is assumed that the graphene film of 1~$\mu$m $ \times$ 1~$\mu$m is under an electric field of 30 kV/cm.  The impurity density is $5 \times 10^{11}$~cm$^{-2}$.  The experimental data from Refs.~~\onlinecite{Meric2008} and \onlinecite{Dorgan2010} are for the case of SiO$_2$ substrate.

\clearpage
\begin{center}
\begin{figure}
\includegraphics[bb=243 514 416 672]{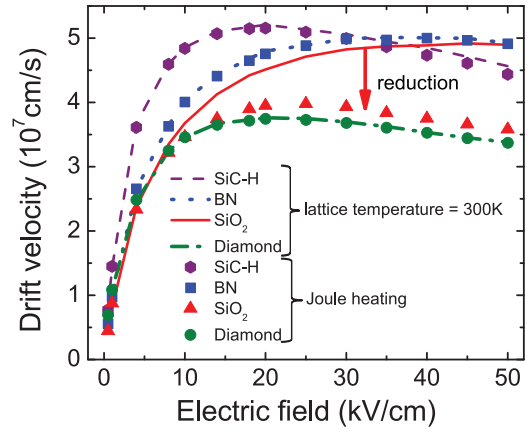}  
\caption{{\large Li {\em et al.}} }
\end{figure}
\end{center}

\clearpage
\begin{center}
\begin{figure}
\includegraphics[bb=243 514 416 672]{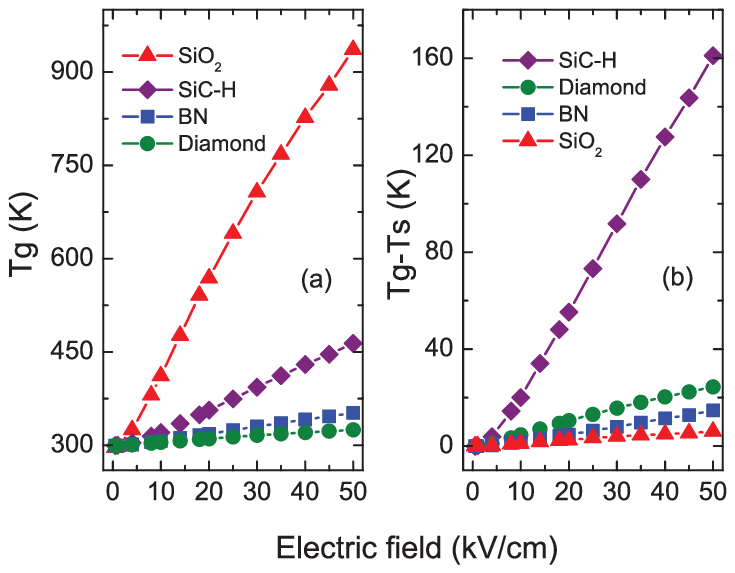}
\caption{{\large Li {\em et al.}} }
\end{figure}
\end{center}

\clearpage
\begin{center}
\begin{figure}
\includegraphics[bb=243 514 416 672]{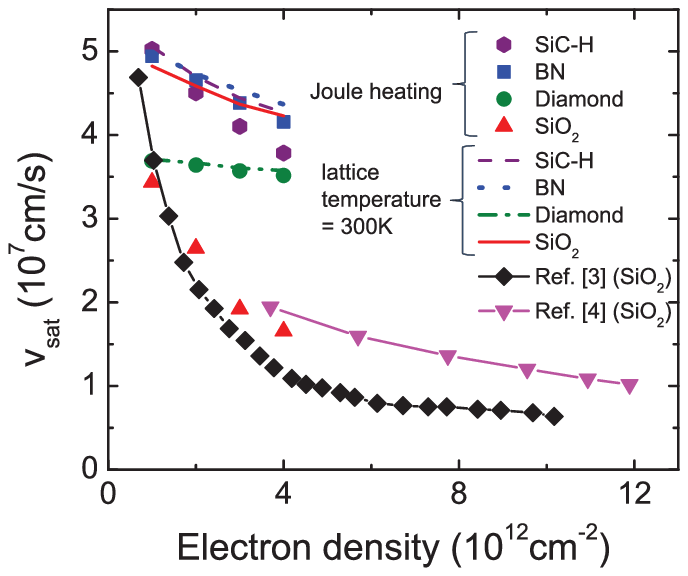}
\caption{{\large Li {\em et al.}} }
\end{figure}
\end{center}

\end{document}